\documentclass{emulateapj}

\usepackage{graphicx}
\usepackage{amssymb}
\usepackage{longtable}
\usepackage{epsf}
\usepackage{apjfonts}
\usepackage{mathptmx}

\newcommand{\myemail}{danilom@astro.yale.edu}

\slugcomment{Received 2007 April 11; accepted 2007 May 24; published  2007 June 21}

\shorttitle{LFs at $z \geq 2$: observations vs. predictions}
\shortauthors{Marchesini \& van Dokkum}

\begin{document}

\title{Assessing the Predictive Power of Galaxy Formation Models:
A Comparison of Predicted and Observed Rest-frame Optical
Luminosity Functions
at $2.0 \leq z \leq 3.3$}

\author{Danilo Marchesini\altaffilmark{1} and 
  Pieter G. van Dokkum\altaffilmark{1}}

\altaffiltext{1}{Department of Astronomy; Yale Center for Astronomy and Astrophysics, Yale University, New Haven, 
CT, USA; \myemail}

\begin{abstract}
Recent galaxy formation models successfully reproduce the local
luminosity function (LF) of galaxies by invoking mechanisms to
suppress star formation in low- and high-mass galaxies. As these
models are optimized to fit the LF at low redshift, a crucial question
is how well they predict the LF at earlier times. Here we compare
recently measured rest-frame $V$-band LFs of galaxies at redshifts
$2.0 \leq z \leq 3.3$ to predictions of semianalytic models by
De Lucia \& Blaizot and Bower et al. and hydrodynamic simulations by
Dav\'e et al.. The models succeed for some luminosity and redshift
ranges and fail for others. A notable success is that the Bower et
al.\ model provides a good match to the observed LF at $z\sim
3$. However, all models predict an increase with time of the
rest-frame $V$-band luminosity density, whereas the observations show
a decrease. The models also have difficulty matching the observed
rest-frame colors of galaxies. In all models the luminosity density of
red galaxies increases sharply from $z\sim3$ to $z\sim2.2$, whereas it
is approximately constant in the observations. Conversely, in the
models the luminosity density of blue galaxies is approximately
constant, whereas it decreases in the observations. These discrepancies
cannot be entirely remedied by changing the treatment of dust and
suggest that current models do not yet provide a complete description
of galaxy formation and evolution since $z\sim 3$.
\end{abstract}


%====================================================================

\section{Introduction}

In the current paradigm of structure formation, dark matter (DM) halos
build up in a hierarchical fashion through the dissipationless
mechanism of gravitationally instability. The assembly of the stellar
content of galaxies is instead governed by much more complicated
physical processes, often dissipative and non-linear, which are
generally poorly understood. To counter this lack of understanding,
prescriptions are employed in the galaxy formation models. One of the
fundamental tools for constraining the physical processes encoded in 
these models is the luminosity function (LF), since its shape retains 
the imprint of galaxy formation and evolution processes.

The faint end of the LF can be matched with a combination of supernova
feedback and the suppression of gas cooling in low-mass halos due to a
background of photoionizing radiation (e.g., \citealt{benson02}).
Matching the bright end of the LF has proven more challenging. Very
recent implementation of active galactic nucleus (AGN) feedback in 
semianalytic models (SAMs) has yielded exceptionally faithful 
reproductions of the observed local rest-frame $B$- and $K$-band global 
LFs (\citealt{bower06}; \citealt{croton06}; see also \citealt{granato04}), 
including good matches to the local rest-frame $B$-band LFs of red and 
blue galaxies (although with some discrepancies for faint red galaxies;
\citealt{croton06}).

The excellent agreement between observations and models at $z\sim0$ is
impressive but is partly due to the fact that the model parameters were
adjusted to obtain the best match to the local universe. A key
question is therefore how well these models predict the LF at earlier
times. The SAMs of \citet{croton06}, \citet{delucia06}, and 
\citet{bower06} have been compared to observations at $0<z<2$ 
\citep[see][]{bower06,kitzbichler06}. Although the agreement is 
generally good, \citet{kitzbichler06} infer that the
abundance of galaxies near the knee of the LF at high redshift is
larger in the SAMs than in the observations (except possibly for the
brightest objects), in an apparent reversal of previous studies (e.g.,
\citealt{cimatti02}). 

Recently, the rest-frame optical LF has been accurately measured in the
redshift range $2.0 \leq z \leq 3.3$, using a combination of the
$K$-selected MUSYC, GOODS, and FIRES surveys \citep{marchesini07}.  In
this Letter we compare the observed LF to that predicted by
theoretical models in this redshift range, in order to test the
predictive power of the latest generation of galaxy formation
models. We also compare the observed LF to predictions from smoothed 
particle hydrodynamics (SPH) simulations, which have so far only been 
compared to data at $z\sim6$ \citep{dave06}. We note that these 
comparisons are effectively the rest-frame equivalent of the test 
proposed by \citet{kauffmann98}. We assume $\Omega_{\rm M}=0.3$, 
$\Omega_{\rm \Lambda}=0.7$, and $H_{\rm 0}=70$~km~s$^{-1}$~Mpc$^{-1}$.  
All magnitudes are in the AB system, while colors are on the Vega system.

%====================================================================

\section{The observed luminosity functions} \label{olf}

The observed rest-frame optical LFs at $z \geq2$ have been taken from
\citet{marchesini07}. Briefly, they presented the galaxy LFs in the
rest-frame $B$-band (at $2.5<z \leq 3.5$ and $2 \leq z \leq 2.5$),
$V$-band (at $2.7 \leq z \leq 3.3$), and $R$-band (at $2 \leq z \leq
2.5$), measured from a $K$-selected sample constructed from the
MUltiwavelength Survey by Yale-Chile (MUSYC; \citealt{quadri06}), the
ultradeep Faint InfraRed Extragalactic Survey (FIRES;
\citealt{franx03}), and the Great Observatories Origins Deep Survey
(GOODS; \citealt{giavalisco04}; Chandra Deep Field--South).  This
$K$-selected sample, comprising a total of $\sim$990 galaxies with
$K_{\rm s}^{\rm tot}<25$ at $2 \leq z \leq 3.5$, is unique for its
combination of surveyed area ($\sim$380~arcmin$^{2}$) and large range
of luminosities.

In this Letter we limit our comparison between observed and predicted
LFs to the rest-frame $V$ band, at the two redshift intervals $2.7
\leq z \leq 3.3$ (directly taken from \citealt{marchesini07}) and $2
\leq z \leq 2.5$ (derived in the same way as described in
\citealt{marchesini07}). The results are qualitatively similar for
other rest-frame bands.

%====================================================================

\section{The model-predicted luminosity functions} \label{plf}

The \citet{bower06} SAM is implemented on the Millennium DM 
simulation described in \citet{springel05}. The details of the 
assumed prescriptions and the specific parameter choices are 
described in \citet{cole00}, \citet{benson03}, and \citet{bower06}.
We have also used the outputs\footnote{Available at
http://www.mpa-garching.mpg.de/Millennium \citep{lemson06}; see also
footnote~6.}  from the SAM of \citet{croton06} as updated by
\citet{delucia06}.  This model differs from the SAM of \citet{bower06}
in many ways. The scheme for building the merger trees is different in
detail, as are many of the prescriptions adopted to model the baryonic
physics, most notably those associated with the growth of and the
feedback from SMBHs in galaxy nuclei and the cooling model (see
\citealt{kauffmann00}; \citealt{springel01}; \citealt{delucia04}; 
\citealt{delucia06} for details). Finally, we have compared the 
observed LFs with the predictions from the cosmological SPH simulations 
of \citet{oppenheimer06}, already used in \citet{finlator06} to 
constrain the physical properties of $z \sim 6$ galaxies. The key 
ingredient of these simulations is the inclusion of superwind feedback, 
critical to avoid the overprediction by the simulations of the 
observed global star formation rate by reducing the reservoir of gas 
available for star formation \citep{springel03}.  Specifically, we 
used the ``momentum-driven wind'' model used in Finlator et al. 
(2007, namely, their ``jvzw'' model; see also \citealt{oppenheimer06} 
for detailed descriptions). We have used the 32~h$^{-1}$~Mpc box 
simulation, combined with a 64~h$^{-1}$~Mpc box to better sample the 
bright end of the LF. We note that the key difference between the 
AGN feedback implementation in the SAMs and the superwind feedback is 
that the former does not require star formation.

Computing the SAM-predicted rest-frame $V$-band LFs is
straightforward, as the catalog is complete in the luminosity range of
interest and has no redshift errors\footnote{To derive the model-predicted 
LF in a specific redshift interval, we averaged the number of galaxies as 
function of rest-frame V-band magnitude (with dust modeling included) of 
all redshift snapshots in the targeted redshift interval.}.  We also 
extracted rest-frame colors of the galaxies in order to determine the LF 
for red and blue galaxies separately.

%====================================================================

\section{Results} \label{results}

\begin{figure}
\epsscale{0.95}
\plotone{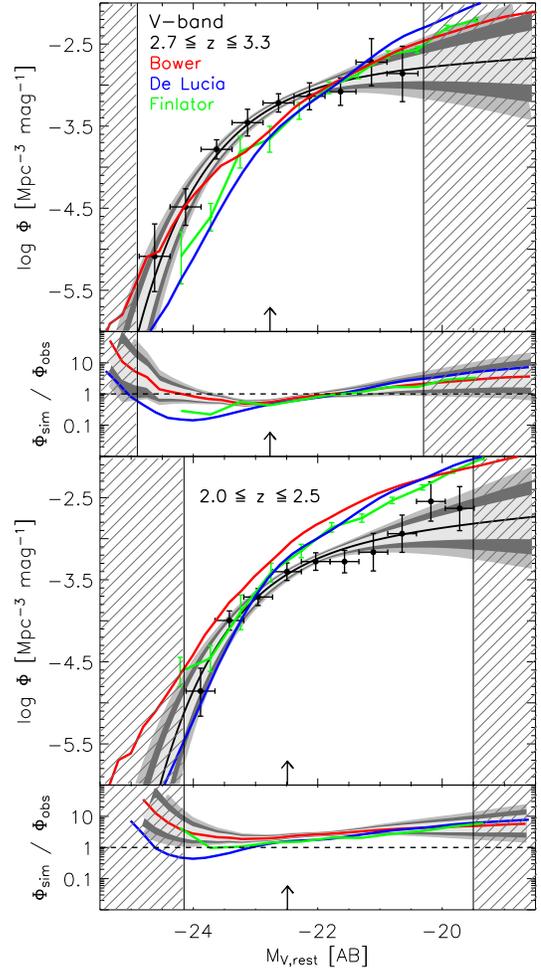}
\caption{Comparison between the rest-frame $V$-band observed global
LFs and those predicted by models. The observed LFs are plotted with
black circles ($1/V_{\rm max}$ method) with 1~$\sigma$ error
bars (including field-to-field variance) and by the black solid line
(maximum likelihood method) with 1, 2, and 3~$\sigma$ solutions
({\it gray shaded regions}). The arrow shows the observed value of
$M^{\star}$. Red lines show predictions from the \citet{bower06} SAM,
blue lines from the \citet{delucia06} SAM, and green lines from the
\citet{finlator06} SPH model. Poisson errors ($1~\sigma$) are shown for
the SPH model only, as they are very small for the SAMs. In the small
panels, the ratio between the predicted and the observed LFs is
plotted, together with the 1, 2, and 3~$\sigma$ errors for the
\citet{bower06} SAM ({\it gray shaded regions}). The oblique line regions 
delimit the comparison to the luminosity range probed by the sample 
of \citet{marchesini07}.
\label{fig1}}
\end{figure}

The comparison between the observed rest-frame $V$-band LFs of all
galaxies at $2.7 \leq z \leq 3.3$ and $2 \leq z \leq 2.5$ with those
predicted by the theoretical models is shown in Figure~\ref{fig1}.  It
is immediately obvious that the models do not yet
provide a precise description of galaxy evolution.  Differences
between the various models, and discrepancies between model
predictions and data, are still as large as a factor of $\sim 5$ for
certain luminosity and redshift ranges.

At $2.7 \leq z \leq 3.3$, the global LF predicted by the SAM of
\citet{bower06} agrees well with the observed LF, although the SAM 
slightly underpredicts the density of galaxies around the knee of the LF. 
However, while at $2 \leq z \leq 2.5$ the shape of the observed LF is 
broadly reproduced by
the SAM, the predicted characteristic density $\Phi^{\star}$ is
$\sim$2.5 times larger than the observed value.  The SAM of
\citet{delucia06} has difficulty with both the normalization and the
slope of the LF, which is too steep.  At $2.7 \leq z \leq 3.3$, the De
Lucia model matches the faint end but underpredicts (by a factor of
$\sim$2--4) the bright end.  At $2 \leq z \leq 2.5$, instead, the
predicted LF matches the bright end but overpredicts the faint end by
a factor of $\gtrsim 2$.  The SPH simulations of \citet{finlator06}
predict LFs that are qualitatively similar to those predicted by the
two SAMs, although the former are characterized by larger
uncertainties, due to the much smaller simulated volume.

We quantified these results by determining the luminosity density
$j_{\rm V}$ (obtained by integrating the LF) for the observations and
models.  The luminosity density is a more robust measure than
$M^{\star}$, $\Phi^{\star}$, and the faint-end slope $\alpha$, because
the errors in these parameters are highly correlated.  The observed
$j_{\rm V}$ ($j^{\rm obs}_{\rm V}$) has been estimated by integrating
the best-fit Schechter function down to $M_{\rm V}=-19.5$, which is
the faintest luminosity probed by the $K$-selected sample\footnote{As
in \citet{marchesini07} the 3~$\sigma$ error on $j_{\rm V}$ was
calculated by deriving the distribution of all the values of $j_{\rm
V}$ within the 3~$\sigma$ solutions of the Schechter LF parameters
from the maximum-likelihood analysis, including in quadrature a 10\%
contribution from photometric redshift uncertainties. Using a brighter 
integration limit of the LF ($M_{\rm V}=-20.4$) does not change 
the results of the comparison significantly.}.  To estimate
$j_{\rm V}$ from the SAM ($j^{\rm SAM}_{\rm V}$), we have fitted the
predicted LFs with a Schechter function, leaving $M^{\star}$,
$\Phi^{\star}$, and $\alpha$ as free parameters, applying the same
limits as to the data.

\begin{figure}
\epsscale{0.95}
\plotone{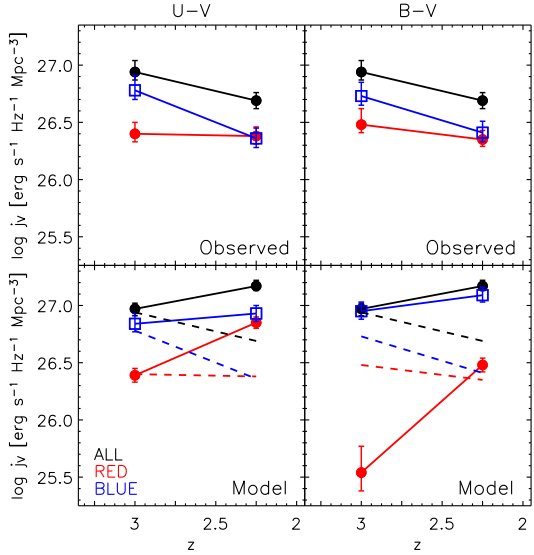}
\caption{{\em Top panels:} Observed luminosity density ($j^{\rm
obs}_{\rm V}$) as function of redshift of all ({\it black circles}),
red ({\it red circles}), and blue ({\it blue squares}) galaxies,
splitting the sample based on rest-frame $U-V$ ({\it left panels}) 
and $B-V$ ({\it right panels}) colors. {\em Bottom panels:} Luminosity 
density predicted by the SAM of \citet{bower06} ($j^{\rm SAM}_{\rm V}$) 
as function of redshift; symbols as in top panels; the observed evolution
of $j_{\rm V}$ is also plotted with dashed lines for comparison.
\label{fig2}}
\end{figure}

The comparison between $j^{\rm obs}_{\rm V}$ and $j^{\rm SAM}_{\rm V}$
of \citet{bower06} is shown in Figure~\ref{fig2} ({\it bottom panels}) by the
black lines and data points. The Bower SAM matches the observed
luminosity density at $z\sim 3$. However, the model does not match the
evolution of $j_{\rm V}$. In the model the luminosity increases with
cosmic time, by a factor of $\sim1.6$ from $z\sim3$ to $2.2$. By
contrast, the observed luminosity density {\em decreases} with time,
by a factor of $\sim1.8$ over the same redshift range. Results for the
De Lucia SAM are similar, but for this model the difference between
observed and predicted density is a strong function of the adopted
faint-end integration limit.

%====================================================================

\section{Colors}

We investigated the cause of the discrepancies by splitting the sample
into blue and red galaxies, using their rest-frame colors.  Here we
focus on the Bower model, as it provides the best match to the shape
of the global LF, and a wide range of rest-frame colors are
available. Interestingly, the results depend strongly on the choice of
color: splitting the sample by $U-V$ color (as done in
\citealt{marchesini07}) produces very different results than splitting
by $B-V$ color.

To define red galaxies, we first use the criterion $U-V \geq$0.25, as
done in \citet{marchesini07}.  As shown in the bottom left panel of
Figure~\ref{fig2}, the Bower model reproduces the densities of red and
blue galaxies at $z\sim 3$ extremely well.  The model overpredicts the
densities of red and blue galaxies at $z\sim 2.2$, although it
predicts the correct ratio between the two (roughly 1:1).

Next, we use the criterion $B-V \geq$0.5.\footnote{For observed
galaxies in the Marchesini et al.\ (2007) sample, $U-V=0.25$ roughly
corresponds to $B-V=0.5$.} As can be seen in the top panels of
Figure~\ref{fig2}, this criterion leads to very similar observed
densities of red and blue galaxies as the $U-V$ criterion. However,
the predicted densities are in severe disagreement with the
observations, particularly at $z\sim 3$ (see Fig.~\ref{fig2}, 
{\it bottom right panel}). The red galaxy density at $z\sim 3$ underpredicts
the observed density by a factor of $\sim8$.  Qualitatively similar
results are obtained when $j^{\rm SAM}_{\rm V}$ from the SAM of
\citet{delucia06} is used in the comparison.\footnote{The De Lucia
model provides $B-V$ colors, but no $U-V$ colors.}

Irrespective of the color criterion that is used, we find that the
predicted {\em evolution} of the red and blue luminosity densities is
in qualitative disagreement with the observed evolution.  In the
observations, the moderate evolution of the luminosity density is
mainly driven by a decrease with cosmic time of the density of blue
galaxies, with the red galaxies evolving much less (see also
\citealt{brammer07}).  By contrast, in the SAMs, the moderate
evolution seen in the global LF is in the opposite sense and
dominated by a strong evolution in the red galaxy population.

%====================================================================

\section{Discussion} \label{disc}

The main results of our comparison between the observed and the
model-predicted rest-frame $V$-band LFs of galaxies at $z \geq 2$ are 
(1) the SAM of \citet{bower06} reproduces well the observed LF at 
$z\sim 3$; (2) the models predict an increase with time of the
rest-frame $V$-band luminosity density, whereas the observations show
a decrease; (3) the models predict strong evolution in the red galaxy
population, whereas in the observations most of the evolution is in
the blue population; (4) the models greatly underpredict the abundance
of galaxies with $B-V \geq0.5$ at $z\sim 3$.

\begin{figure*}
\epsscale{0.95}
\plotone{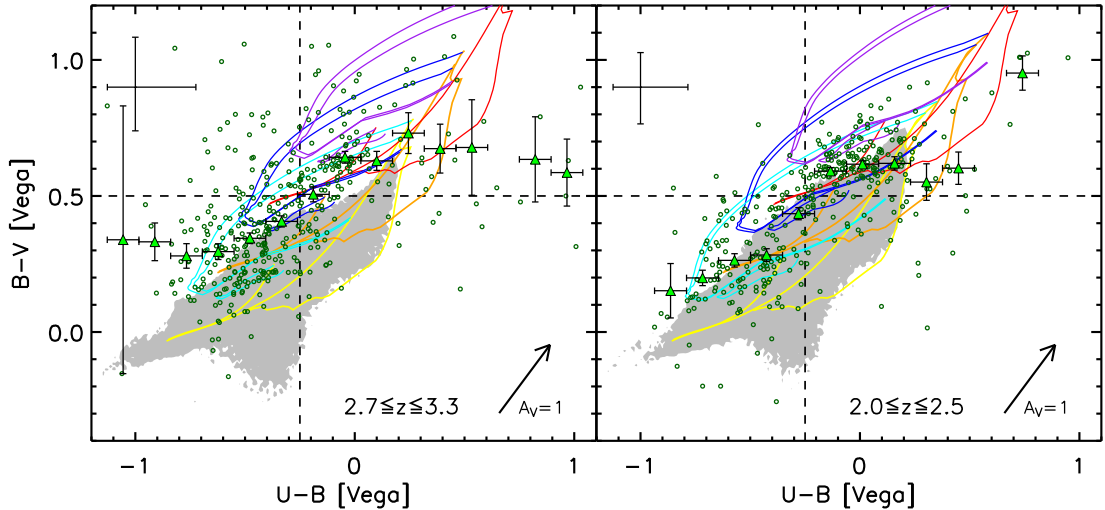}
\caption{$B-V$ vs $U-B$ comparison between observations ({\it open circles}) 
and predictions from the SAM of Bower et al. (2006, {\it gray shaded regions}) 
in the two targeted redshift intervals. The error bars in the
top left corner represent the median errors on the observed
colors. The filled triangles with error bars represent the mean colors
of the observed galaxies and the error on the mean. The yellow,
orange, and red lines show the evolutionary tracks described in
\S~\ref {disc} (CSF, $\tau$300, and SSP models, from top to bottom)
with $A_{\rm V}$=0, 1, and 2, respectively.  The tracks are plotted
from 50~Myr to the age of the universe at the lower limit of the
targeted redshift range. The cyan, blue, and purple lines show the
evolution of the colors after a burst of star formation, for the three
values of $A_{\rm V}$, respectively. The arrow indicates the
extinction vector for $A_{\rm V}$=1. The dashed lines correspond to
$B-V$=0.5 and $U-B=(U-V)-(B-V)=0.25-0.5=-0.25$. Observed galaxies have
redder $B-V$ colors than predicted, possibly due to additional dust
and/or secondary star bursts.
\label{fig3}}
\end{figure*}

The different results obtained for $U-V$ and $B-V$ colors are
interesting, as they may hint at possible ways to improve the models.
We further investigate the disagreement between observed and predicted
colors in the SAM of \citet{bower06} in Figure~\ref{fig3}, which shows
the comparison of observations and predictions in the $B-V$ versus
$U-B$ diagram. While the SAM seems to broadly reproduce the observed
$U-B$ distribution, it predicts galaxies that are systematically bluer
in $B-V$ than the observed galaxies.  We have plotted
evolutionary tracks of stellar population synthesis models constructed
with the \citet{bruzual03} code, assuming three different
prescriptions for the star formation history (SFH): a constant SFH
(CSF), an exponentially declining in time SFH characterized by the
parameter $\tau=300$~Myr ($\tau$300), and an instantaneous burst model
(SSP). We selected the ``Padova 1994'' evolutionary tracks, solar
metallicity, the \citet{chabrier03} initial mass function with lower
and upper mass cutoffs 0.1~M$_{\sun}$ and 100~M$_{\sun}$, and modeled
the extinction by dust using the attenuation law of
\citet{calzetti00}. A new burst of star formation lasting 100~Myr and
contributing 20\% to the mass is also added at $t=2.1 \times
10^{9}$~yr ($t=2.9 \times 10^{9}$~yr) at $z\sim3$ ($z\sim2.2$) to
explore more complex SFHs.

As can be deduced from Fig.~\ref{fig3}, the differences between
observed and predicted colors could be due to larger amount of dust
and/or to more complex SFHs in the observed galaxies. The ad hoc
treatment of dust absorption is a significant and well-known source of
uncertainty in the models.  Modifications to the specific dust model
could partly resolve the differences between observations and SAM
predictions. By simply multiplying the $A_{\rm V}$ in the SAM by a
fixed factor, we were able to better reproduce the observed LFs at
$z\sim2.2$ (although making the faint-end slope of the red galaxy LFs
quantitatively too steep) and to have a better agreement between
observed and predicted colors. However, at $z\sim3$ this simple remedy
is not able to solve the disagreement between the predicted and the 
observed number of $B-V \geq 0.5$ galaxies. We conclude that, while ad 
hoc modifications of the dust treatment might help to alleviate some 
of the found disagreements, it does not seem to be sufficient to 
accommodate the problem with global colors at $z\sim3$.

While our ability to simulate galaxy formation has greatly improved in
the past few years, our results imply that the present understanding
of the physical processes at work in galaxy formation and evolution is
still far from being satisfactory. On the observational side, more
accurate redshift and color estimates would benefit studies of this
kind.  The lack of spectroscopic redshifts is particularly worrying,
as systematic errors in redshift will lead to systematic errors in
colors and luminosities (e.g., \citealt{kriek06}; M. Kriek et al. 2007, 
in preparation).

%====================================================================

\acknowledgments

We are grateful to G.\ De Lucia (the referee) and G.\ Lemson for 
assistance with the Millennium Simulation database and helpful 
clarifications. We thank K.\ Finlator and R.\ Dav\'e for making 
available their SPH simulations, and R.\ Bower for his help with 
obtaining the Bower et al.\ (2006) predictions. D.M. is supported 
by NASA LTSA NNG04GE12G. The authors acknowledge support from 
NSF CARRER AST~04-49678.  The Millennium Simulation databases used 
in this Letter and the Web application providing online access to 
them were constructed as part of the activities of the German 
Astrophysical Virtual Observatory.

%====================================================================


\begin{thebibliography}{}

\bibitem[Benson et al. (2003)]{benson03} Benson, A. J., Bower, R. G., Frenk, C. S., Lacey, C. G., Baugh, C. M., \& Cole, S. 2003, \apj, 599, 38
\bibitem[Benson et al. (2002)]{benson02} Benson, A. J., Lacey, C. G., Baugh, C. M., Cole, S., Frenk, C. S. 2002, \mnras, 333, 156
\bibitem[Bower et al. (2006)]{bower06} Bower, R. G., Benson, A. J., Malbon, R., Helly, J. C., Frenk, C. S., Baugh, C. M., Cole, S., \& Lacey, C. G. 2006, \mnras, 370, 654
\bibitem[Brammer \& van Dokkum (2007)]{brammer07} Brammer, G. B., \& van Dokkum, P. 2007, \apj, 654, L107
\bibitem[Bruzual \& Charlot (2003)]{bruzual03} Bruzual, G., \& Charlot, S. 2003, \mnras, 344, 1000
\bibitem[Calzetti et al. (2000)]{calzetti00} Calzetti, D., Armus, L., Bohlin, R.C., Kinney, A. L., Koornneed, J., \& Storchi-Bergmann, T. 2000, \apj, 533, 682
\bibitem[Chabrier (2003)]{chabrier03} Chabrier, G. 2003, \pasp, 115, 763
\bibitem[Cimatti et al. (2002)]{cimatti02} Cimatti, A., et al. 2002, \aap, 391, L1
\bibitem[Cole et al. (2000)]{cole00} Cole, S., Lacey, C. G., Baugh, C. M., \& Frenk, C. S. 2000, \mnras, 319,168
\bibitem[Croton et al. (2006)]{croton06} Croton, D.~J., et al. 2006, \mnras, 365, 11
\bibitem[Dav\'e et al. (2006)]{dave06} Dav\'e, R., Finlator, K., \& Oppenheimer, D. 2006, \mnras, 370, 273
\bibitem[De Lucia \& Blaizot (2007)]{delucia06} De Lucia, G., \& Blaizot, J. 2007, \mnras, 375, 2
\bibitem[De Lucia et al. (2004)]{delucia04} De Lucia, G., Kauffmann, G., \& White, S. D. M. 2004, \mnras, 349, 1101
\bibitem[Finlator et al. (2007)]{finlator06} Finlator, K., Dav\'e, R., \& Oppenheimer, B. D. 2007, \mnras ~in press
\bibitem[Franx et al. (2003)]{franx03} Franx, M., et al. 2003, \apj, 587, L79
\bibitem[Giavalisco et al.(2004)]{giavalisco04} Giavalisco, M., et al. 2004, \apj, 600, L93
\bibitem[Granato et al.(2004)]{granato04} Granato, G.L., De Zotti, G., Silva, L., Bressan, A., \& Danese, L. 2004, \apj, 600, 580
\bibitem[Kauffmann \& Charlot (1998)]{kauffmann98} Kauffmann, G., \& Charlot, S. 1998, \mnras, 297, L23
\bibitem[Kauffmann \& Haehnelt (2000)]{kauffmann00} Kauffmann, G., \& Haehnelt, M. 2000, \mnras, 311, 576
\bibitem[Kitzbichler \& White (2007)]{kitzbichler06} Kitzbichler, M. G., \& White, S. D. M. 2007, \mnras, 376, 2
\bibitem[Kriek et al. (2006)]{kriek06} Kriek, M., et al. 2006, \apj, 649, L71
\bibitem[Lemson (2006)]{lemson06} Lemson, G., et al. 2006, preprint (astro-ph/0608019)
\bibitem[Marchesini et al.(2007)]{marchesini07} Marchesini, D. et al. 2007, \apj, 656, 42
\bibitem[Oppenheimer \& Dav\'e (2006)]{oppenheimer06} Oppenheimer, B. D., \& Dav\'e, R. 2006, \mnras, 373, 1265
\bibitem[Quadri et al. (2007)]{quadri06} Quadri, R., et al. 2007, \aj, in press (astro-ph/0612612)
\bibitem[Springel \& Hernquist (2003)]{springel03} Springel, V., \& Hernquist, L. 2003, \mnras, 339, 312
\bibitem[Springel et al. (2001)]{springel01} Springel, V., White, S. D. M., Tormen, G., \& Kauffmann, G. 2001, \mnras, 328, 726
\bibitem[Springel et al. (2005)]{springel05} Springel, V., et al. 2005, \nat, 435, 629
\end{thebibliography}
\end{document}